\documentstyle[iopconf1,epsfig,,amssymb]{article}
\begin{document}
\title{Atmospheric and Astrophysics Neutrinos with MACRO}

\author{Teresa Montaruli\dag\footnote{E-mail:
montaruli@ba.infn.it} for the MACRO Collaboration}

\affil{\dag\ INFN, Sezione di Bari, Via Amendola 173, 70126, Italy}
%

\beginabstract
MACRO can detect three topologies of neutrino induced events, corresponding to 
different parent neutrino energies. The most numerous sample is made of  
upward throughgoing muons induced by atmospheric neutrinos of average energy 
100 GeV. 
The ratio of the observed over the expected events is $0.74 \pm 0.036_{stat} 
\pm 0.046_{sys} \pm 0.13_{theor}$.
The observed zenith distribution does not fit the expected one in 
the no-oscillation
hypothesis, giving a maximum $\chi^2$ probability of 0.1$\%$. Considering the 
$\nu_{\mu}-\nu_{\tau}$ oscillation hypothesis the best 
probability (17$\%$) is obtained for 
maximum mixing and $\Delta m^{2}$ of a few times 10$^{-3}$ eV$^{2}$.
The other detected samples are due to 
internally produced events and upward-going
stopping muons, corresponding to an average neutrino energy of around 4 GeV.
The results concerning these lower energy sample show a deficit and a shape
of the angular distributions in agreement with that 
predicted by the oscillation
model suggested by the higher energy sample.

No evidence of a neutrino signal due to dark matter particles 
in the direction of the core of the 
Earth and of the Sun has been found among the background due to atmospheric
neutrinos and limits have been set. 
The neutralino hypothesis is investigated and limits on its mass are given.
\endabstract

\section{Introduction}

In the past years, the water Cherenkov experiments IMB and Kamiokande have
observed around the 60$\%$ of the expected double ratio of 
contained muon neutrino to electron neutrino interactions 
\cite{IMBKam}. More recently, the Soudan 2
experiment has confirmed the anomaly using an iron--based detector 
\cite{Soudan2}. 
On the other hand, the earlier results from the iron 
calorimeters NUSEX and Frejus are consistent with the expected number 
of contained events though with smaller statistics \cite{NUSFre}.
 
Here we present the measurement of the high energy muon neutrino flux and its 
interpretation in terms of neutrino oscillations and the 
first results concerning the low
energy events in MACRO.

\section{Neutrino detection with MACRO}

The MACRO detector important features for neutrino 
detection are its large area,
fine tracking granularity, symmetric electronics 
with respect to upgoing versus downgoing
particles, good timing and angular resolutions. 
Moreover, the rock overburden with 
minimum thickness of 3150 hg/cm$^2$, 
significantly larger than for other experiments
(e.g. Baksan and IMB), reduces the surface muon flux of a factor
of around 10$^6$, hence providing a shield to 
possible sources of background induced
by downgoing atmospheric muons. These features permit a fully--automated 
analysis of both downgoing and upgoing muons. The symmetric time measurement 
provides an important tool
to study the detector acceptance and possible sources of background
using the huge statistics of downward-going muons. 

The MACRO detector, described in Ref.~\cite{MACRO93} 
is located in the Hall B of the Gran Sasso Laboratory. 
It is a large rectangular box of $76.6 \times
12 \times 9.3$ m$^3$, divided longitudinally in 
six similar supermodules and vertically
in a lower part (4.8 m high) and an upper part (4.5 m high), called ``attico''.
The lower half of the detector is filled with rock absorber between
horizontal streamer tube planes, while the attico is empty and contains the
electronics racks and work areas.
There are 10 horizontal streamer planes in the bottom 
half of the detector and 4 planes on the top 
made of 3 cm wire cells and 27$^{\circ}$ stereo strip readouts. 
Six vertical
planes of streamer tubes cover each lateral side of the detector.
The scintillator system consists of around 
600 tons of liquid scintillator inside 12 m long counters which make three
horizontal layers and a vertical layer for each lateral wall.
The time (position) resolution for muons along a scintillator 
box is about 500 ps ($\sim 11$ cm).

\begin{figure}
\begin{tabular}{cc}
\epsfig{figure=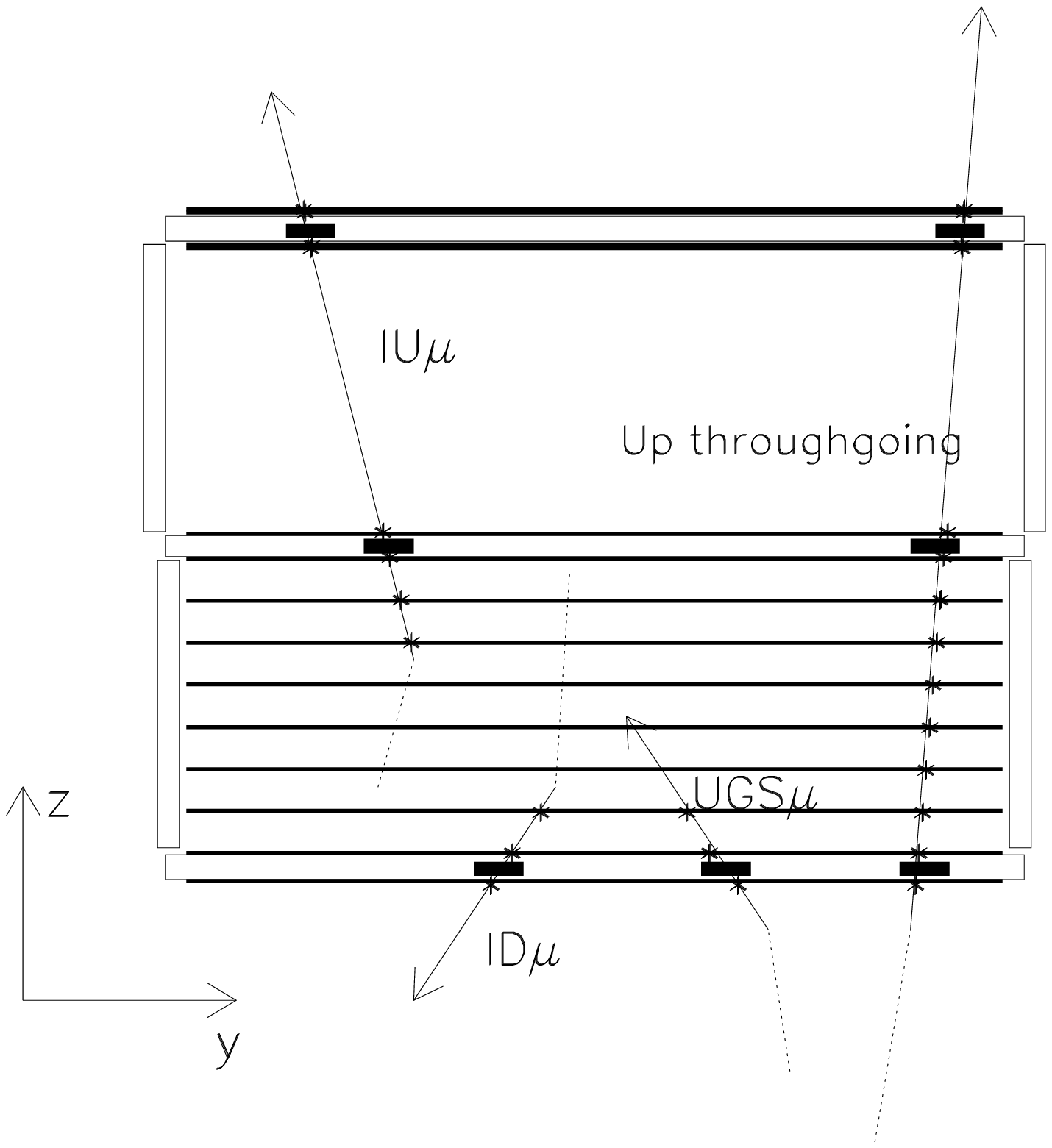,height=6.5cm,width=5.cm}
&\epsfig{figure=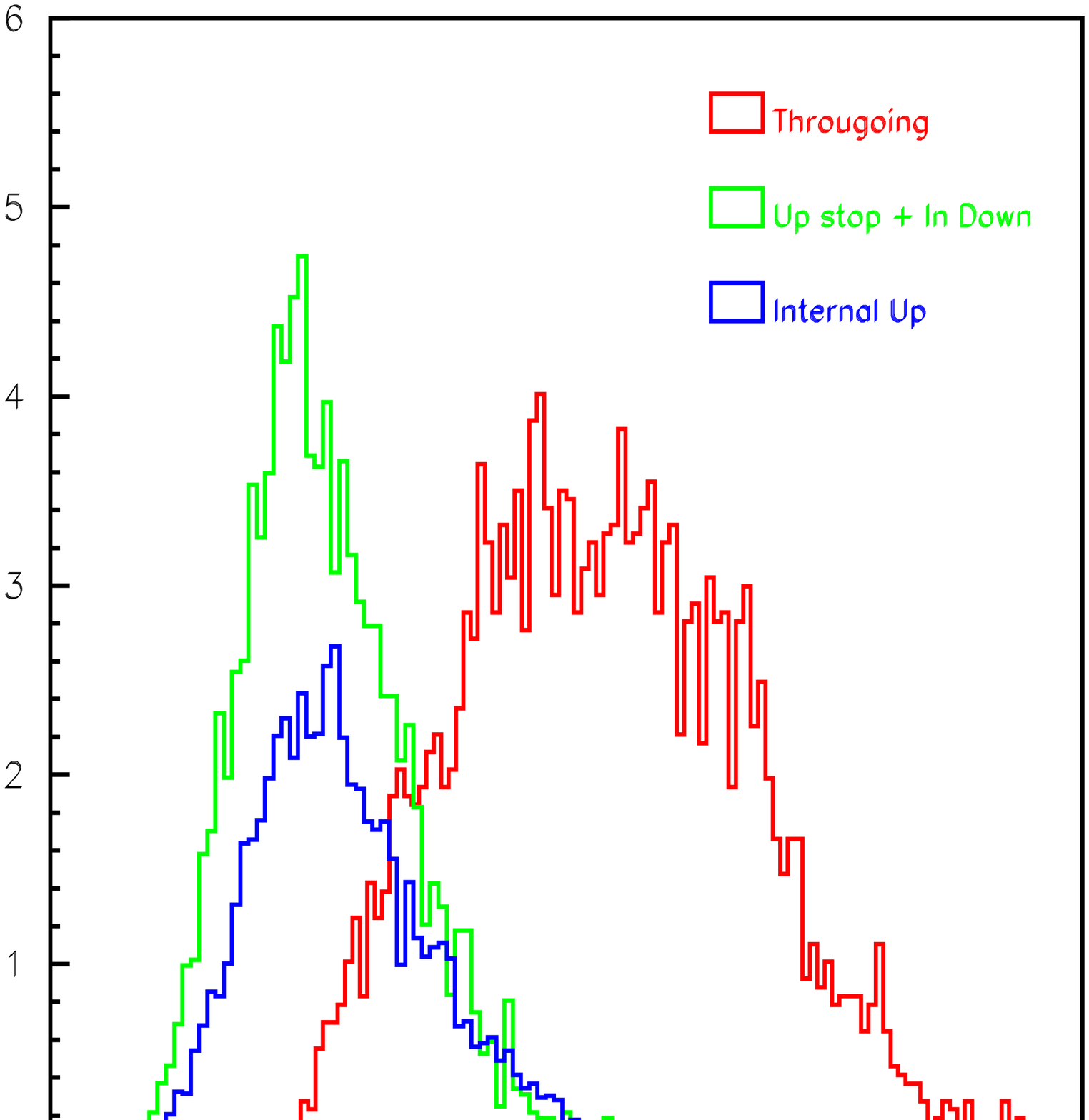,height=5.cm,
width=5.cm}
\end{tabular}
\caption{\label{fig1} On the left: different topologies 
induced by $\nu$ interactions in or around MACRO: {\it Up Through}, 
{\it Internal Up} (IU), {\it Upgoing Stopping} (UGS) $\mu$s and 
{\it Internal Down} (ID).
On the right: distributions of the parent 
$\nu$ energy giving rise to the 3
different $\nu$ event topologies  
computed by Monte Carlo using the same cuts applied to real data and 
normalizing to 1 yr of data taking. 
The average neutrino energies of the samples are: 
$\sim 100$ GeV for {\it Up Through}, $\sim 4$ GeV for  
{\it IU} and $\sim 4$ GeV for 
{\it UGS} $\mu$s and {\it ID}.}
\end{figure}

The schematic plot in Fig.~\ref{fig1} illustrates the three detectable 
neutrino event topologies: {\it Up Through}, {\it Internal Up} (IU), 
{\it Upgoing Stopping} $\mu$s (UGS) and {\it Internal Down } (ID) events. 
The {\it Up Through} and {\it Internal Up} events are measured using the 
time--of--flight tecnique as the events cross at least two scintillator 
layers. 
The {\it Upgoing Stopping} $\mu$s and {\it Internal Down} events cross 
only one scintillator. The lack of time information prevents to distinguish
the two sub-samples.
Fig.~\ref{fig1} shows also the parent neutrino 
energy distributions for the detectable event 
topologies. The average neutrino energies are 
100 GeV for {\it Up Through} muons, 
4 GeV for {\it Upgoing Stopping} $\mu$s together with 
{\it Internal Down} events and the same for {\it Internal Up} events.
The {\it Up Through} $\mu$s and the 
{\it Upgoing Stopping} $\mu$s are externally produced
events coming from $\nu_{\mu}$ interactions 
in the rock surrounding MACRO. In the case of {\it Up Through}, 
the muons cross the detector. 
The absorber in the lower half of the detector 
determines an energy threshold of 1 GeV for vertical muons.
The {\it Internal Up} and {\it Internal Down} 
events come from $\nu$ interactions with
vertex inside the bottom half of the apparatus. 

If the atmospheric neutrino anomaly is the result of 
$\nu_{\mu}-\nu_{\tau}$ oscillations
with $\Delta m^{2} \sim 10^{-3}-10^{-2}$ eV$^2$ it is expected 
a reduction in the flux of the upward throughgoing muons and
a distortion in the shape of their angular distribution. In fact, 
the longer pathlength of neutrinos coming from the 
nadir with respect to that coming from the 
horizon should cause a stronger reduction of the flux near the 
vertical than near the horizontal.
On the other hand, for lower energy events and with these values 
of $\Delta m^{2}$, it is expected a reduction of their flux of about a factor 
two and no distortion in their angular distribution.

The measurements of the three topologies corresponding 
to different energy ranges permit the investigation of 
distinct regions of the oscillation parameter space. 
Moreover, measuring two topologies with energy spectra in the 
same energy range, will allow MACRO to evaluate a double ratio, 
with consequent cancellation of most of the theoretical errors.

\subsection{Upward Throughgoing muons}
\label{Upth}

The signature for muon neutrino induced events is the versus of flight 
of muons: neutrinos can cross amounts of matter from 10 to 10$^{4}$ km 
(the diameter of the Earth) and can be detected through their charged 
current interactions in the surrounding rock as upward-going muons. 
On the other hand, the neutrino induced downward-going muons cannot be
discriminated among the atmospheric muons of many orders of 
magnitude more numerous.

The data have been collected during three periods with different 
detector configurations:
in the first two periods (March 1989 -- November 1991 
and December 1992 -- June 1993)
only lower parts of MACRO were into acquisition; 
during the last period (April 1994 -- November 1997) MACRO was put 
into acquisition in its final configuration including the attico. 
The analysis of the first two samples 
of data is described in Ref.~\cite{upmu95}.
Around 3 years of data have been analyzed with the final configuration 
\cite{upmu98}. 

\begin{figure}
\begin{tabular}{cc}
\epsfig{file=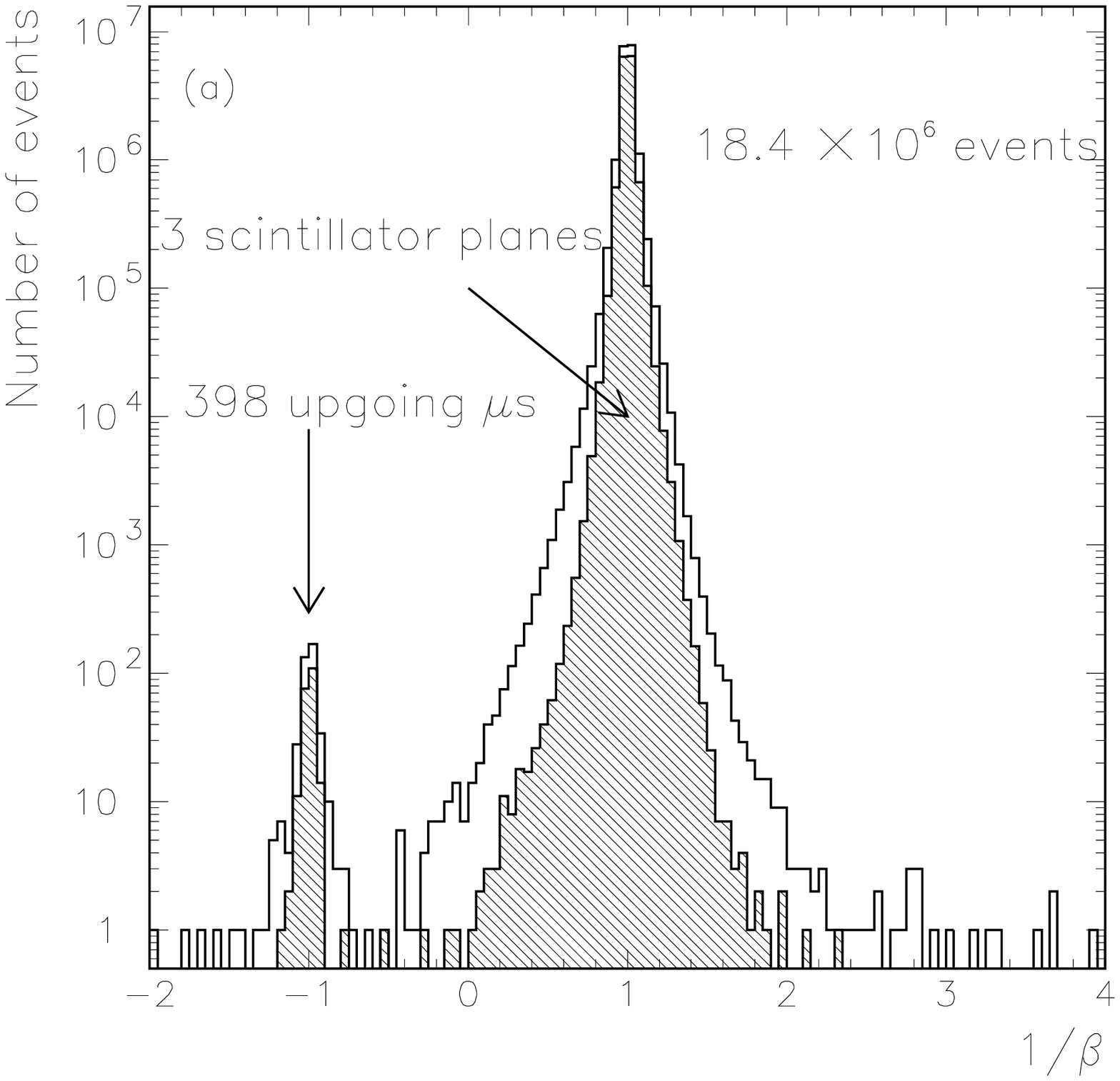,height=6.cm,width=5.cm}
&\epsfig{file=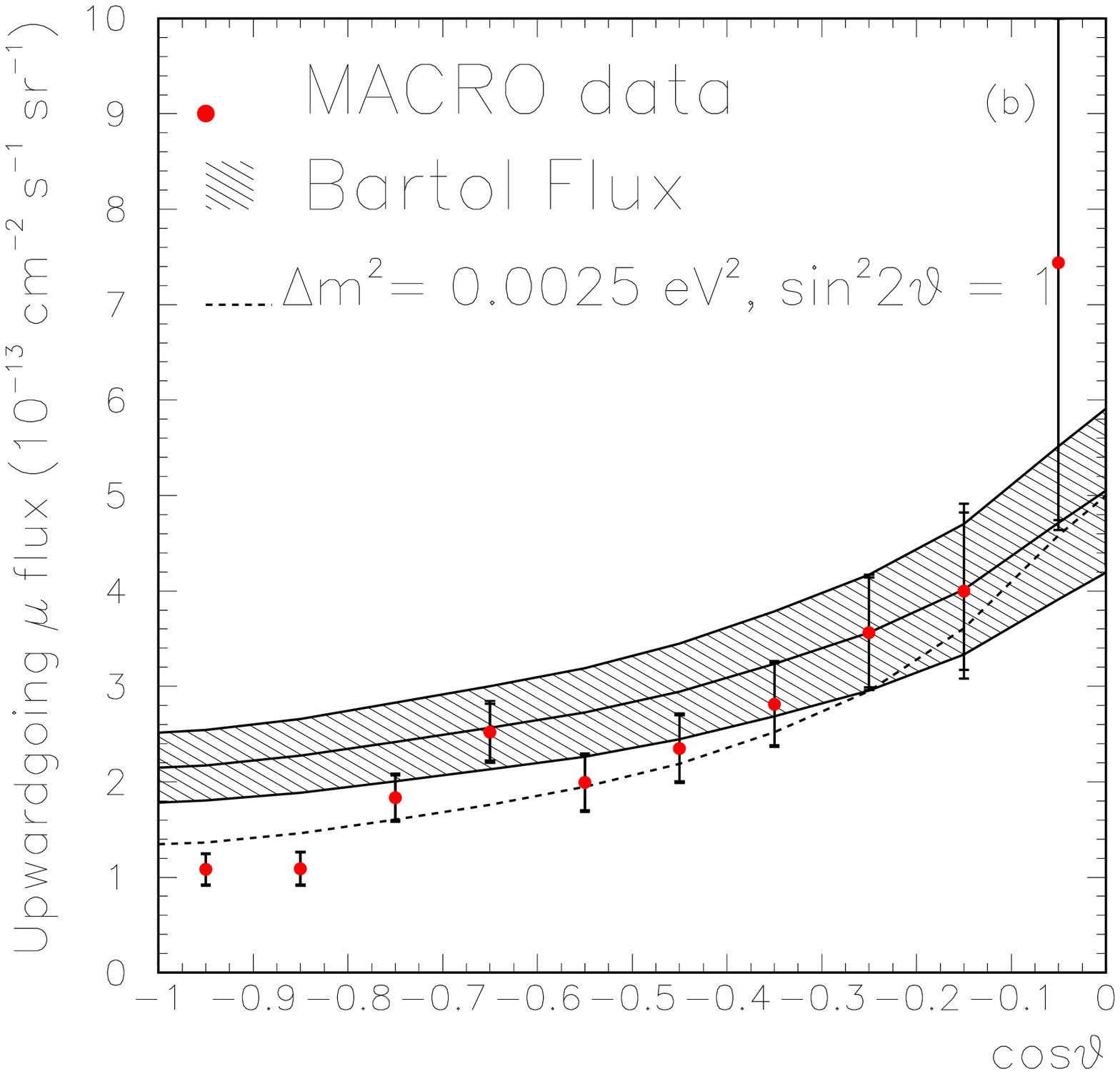,height=6.cm,
width=6.cm}
\end{tabular}
\caption{\label{fig2} (a) $1/\beta$ distribution for the full detector data. 
The downgoing muon peak contains $\sim 18.4$ millions 
of events and the well separated upward throughgoing muon peak contains 398
events. 
The shaded part of the distribution is for the  
events which cross 3 scintillator layers.
(b) Zenith distribution of the flux of upward throughgoing $\mu$s with
$E_{\mu} > 1$ GeV for the data and the Monte Carlo. 
The solid line shows the expectation for no-oscillations with 
the 17$\%$ theoretical uncertainty (shaded region). The dashed
line is the prediction for an oscillated flux with $\sin^{2}2\theta = 1$
and $\Delta m^{2} = 0.0025$ eV$^{2}$.}
\end{figure}

In Fig.~\ref{fig2}(a) the $1/\beta$ distribution is shown ($1/\beta$ 
is proportional to the measured time--of--flight). 
Muons downward-going through the detector are expected 
to have $1/\beta \sim + 1$,
while muons moving upward are expected to have $1/\beta \sim - 1$.
This distribution is obtained by imposing several cuts to remove the
background caused by radioactivity in narrow coincidence with muons, 
showering events and multiple muons which may result in bad time 
reconstruction.
The most important requirement is that the position along 
the scintillator counter measured using the streamer tubes has to differ from
the one from the times measured at the two ends of the counter by less than 
$\pm 70$ cm ($\pm 140$ cm for nearly horizontal tracks). 
When a muon crosses three scintillator layers (almost 50$\%$ of the tracks), 
there is redundancy in the time measurement and $1/\beta$ is calculated 
by a linear fit of the times as a function of the pathlength. 
Tracks with poor fits are rejected. 
Other minor cuts are applied for tracks which intercept only 
two scintillator layers. 
The achieved rejection factor of upward throughgoing muons 
with respect to atmospheric muons is $\sim 10^{-7}$.

The requirement that throughgoing muons should cross at least 
200 g/cm$^{2}$ has been applied to reduce at the percent level the 
background due to an undetected downward-going muon which interacts 
in the rock surrounding the detector and produces a detected 
upgoing particle (in most of the cases a pion) \cite{Spurio}.
Finally, we have observed that a large number of nearly horizontal upgoing 
muons ($\cos\theta \ge -0.1$) come from the azimuth angle region between
$-30^{\circ} to 120^{\circ}$, which corresponds to a cliff in the direction
of Teramo. We exclude this angular region due to the unsufficient rock
coverage.
 
The total number of upward throughgoing muons with $1/\beta$ in the interval
$[-1.25, -0.75]$ is 479. Based on 
events outside the upward-going muon 
peak, we estimate a background of $9 \pm 5$
events due to incorrect $\beta$ measurement 
(such as in the case of more tracks crossing the same
scintillator counter)
and that there are $8 \pm 3$ events 
resulting from upward-going charged particles produced by downward-going 
muons in the rock. Finally, it is estimated
that $11 \pm 4$ events are the result of $\nu$ interactions in the 
bottom scintillator layer which satisfy the upward-going 
muon analysis requirements. Hence, removing the backgrounds and the internally
produced estimated events, the total number of measured upward throughgoing
muons is 451. 

\begin{figure}
\begin{tabular}{cc}
\epsfig{file=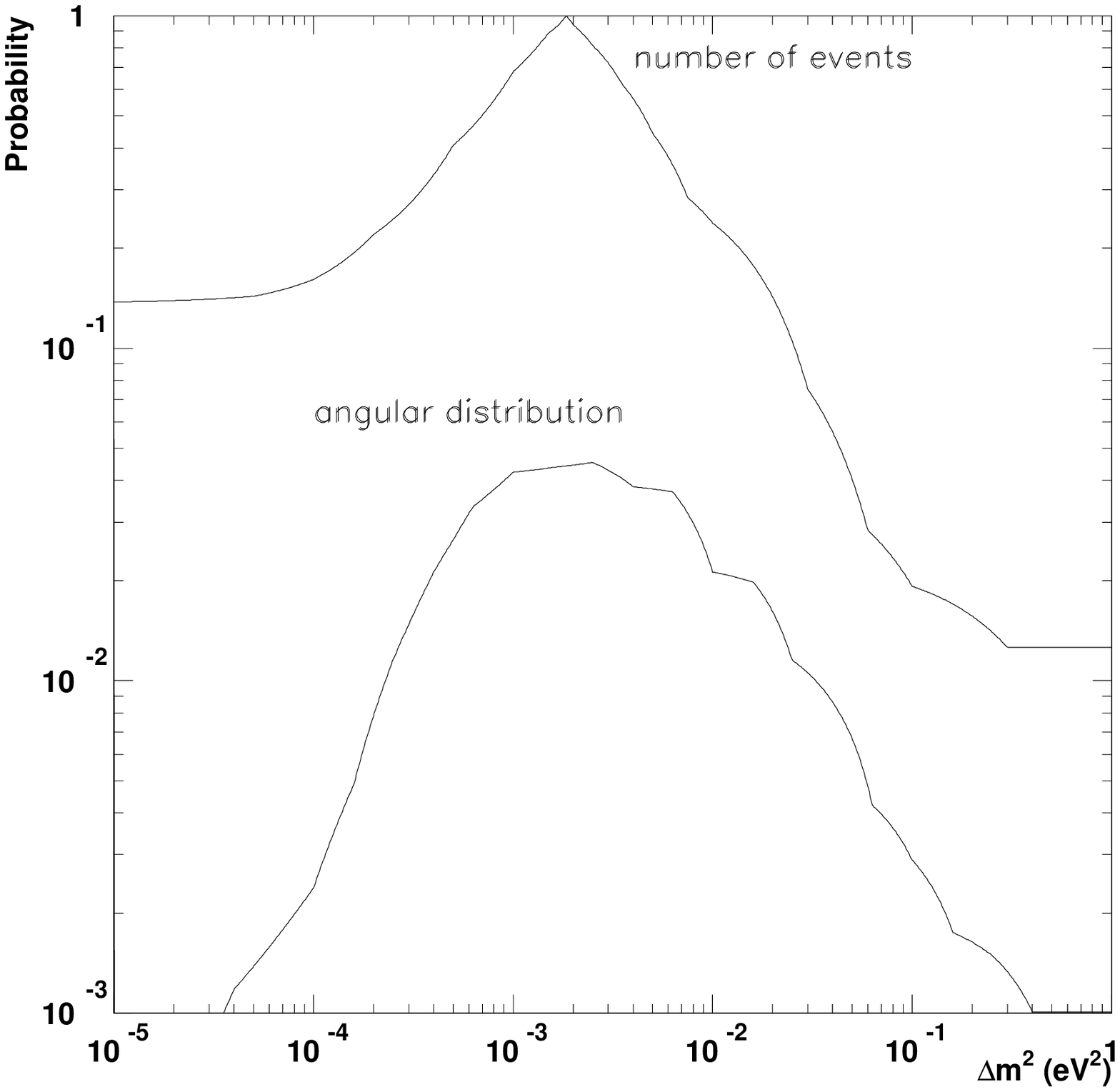,height=5.5cm,
width=4.5cm}&
\epsfig{file=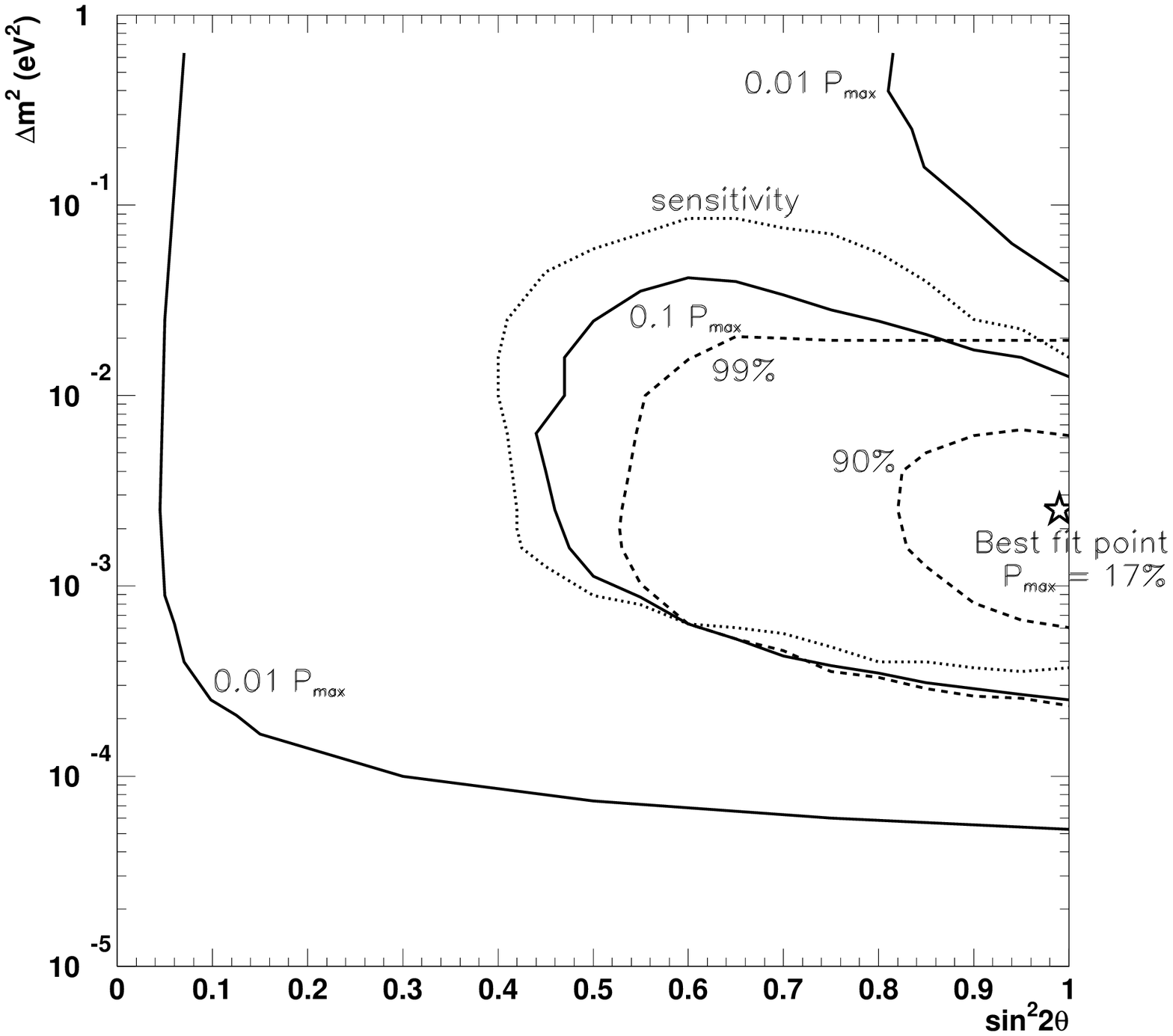,height=5.5cm,
width=5.cm}
\end{tabular}
\caption{\label{fig3} On the left: probabilities for 
obtaining the observed MACRO
zenith distribution shape and total
number of events in the $\nu_{\mu} \rightarrow \nu_{\tau}$ oscillation 
hypothesis as a function of $\Delta m^{2}$ and for $\sin^{2}2\theta = 1$.
On the right: probability contours from the combined probabilities in the
plot on the left. 
The best probability is 17$\%$ and the 
iso-probability contours (solid lines) 
are shown for 10$\%$ and 1$\%$ of this value. 
The confidence regions (90$\%$ and $99\%$ c.l.) calculated according to 
Ref.~\protect\cite{Feldman} (dashed lines) and 
the sensitivity curve of the experiment (dotted line) are shown.}
\end{figure}

The expected upward-going muon 
flux is obtained using the Bartol $\nu$ flux \cite{Agrawal}, the
Morfin and Tung parton distributions set $S_{1}$ \cite{Morfin} for the
calculation of the $\nu$ CC cross-sections (chosen because of the good 
agreement of the total cross section with the
world average at $E_{\nu} = 100$ GeV). The propagation of muons to the
detector has been done using the energy loss calculation by Lohmann \etal
\cite{Lohmann}.  
The total theoretical uncertainty in the calculation of the upward-going muon 
flux, obtained summing in quadrature the errors is $\pm 17\%$.
This error concerns the normalization of the flux, not the shape which does 
not vary changing the inputs of the calculation and it is known at the
level of few percents \cite{Lipari}.
The total number of expected events is 612, giving a ratio of the 
observed to the expected events of $0.74 \pm 0.036_{stat} \pm 0.046_{sys} 
\pm 0.13_{theor}$.
Fig.~\ref{fig2}(b) shows the zenith distribution of the measured flux
of upward throughgoing muons with $E_{\mu} > 1$ GeV compared to the
expected one. The dashed line represents the flux obtained in the hypothesis
of $\nu_{\mu} \rightarrow \nu_{\tau}$ oscillation with maximum mixing and 
$\Delta m^{2} = 0.0025$ eV$^{2}$. 
The $\chi^{2}$ test applied to the angular
distribution gives $\chi^{2} = 26.1$ for 8 d.o.f. (probability of
0.1$\%$ for a shape at least this different from the expectation) 
for the no-oscillation hypothesis excluding the last bin due to a possible
contamination of atmospheric horizontal muons and normalizing the prediction
to the data (the main theoretical error concerns the normalization).  
The best $\chi^{2}$ when testing the $\nu_{\mu} \rightarrow \nu_{\tau}$
hypothesis is 15.8 for $\Delta m^{2}$ around 0.0025 eV$^{2}$ and maximum
mixing (if the bound on the meaningful interval of $sin^{2}2\theta$ 
is released the best $\chi^{2}$ is found for mixing greater than 1).
The independent probabilities for obtaining the observed number 
of events and the angular shape of the distribution calculated for 
maximum mixing and varying $\Delta m^{2}$ are shown in
Fig.~\ref{fig3} (plot on the left). The $\Delta m^{2}$ region
in which the two probabilities have their maximum is similar.
In Fig.~\ref{fig3} (plot on the right) 
the solid lines represent the iso-probability contours
in the oscillation parameter space obtained combining the two 
independent probabilities.
The curves correspond to 10$\%$ and 1$\%$ of the maximum probability of 17$\%$
obtained for maximum mixing and $\Delta m^{2} \sim 0.0025$ eV$^{2}$. 
The dashed lines are the contours of the confidence regions at the 90$\%$
and $99\%$ c.l. based on the Monte Carlo prescription in Ref.~\cite{Feldman}.
This prescription assumes that the hypothesis is correct.
The sensitivity curve results from the preceding prescription if the data
and Monte Carlo happened to be in perfect agreement at the maximum probability
point. The allowed regions are smaller than the sensitivity curve due to the
fact that the absolute minimum $\chi^{2}$ is outside the physical region.
The same procedure has been applied considering the hypothesis of muon 
neutrino oscillations into a sterile neutrino \cite{Smirnov} and the maximum 
probability obtained is 2$\%$.   
\begin{figure}
\epsfig{file=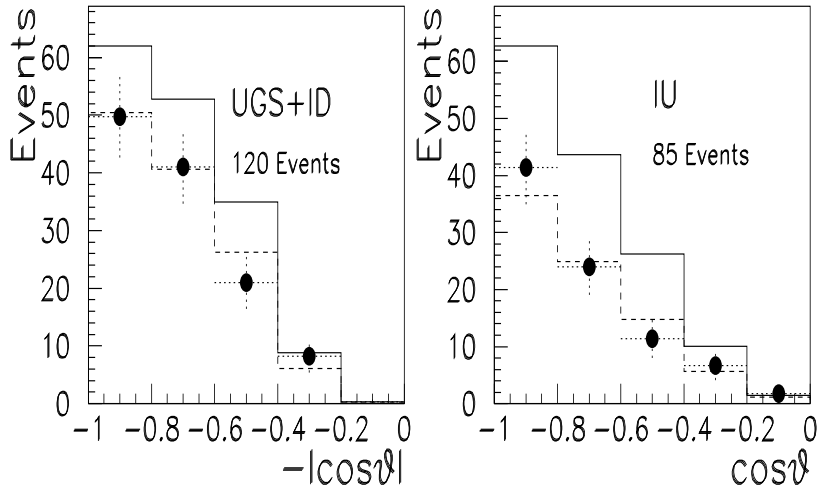,height=5.5cm,
width=10.cm}
\caption{\label{fig4} Comparison between measured and expected low energy 
event distributions vs $\cos\theta$. The dashed line is obtained assuming
$\nu_{\mu} \rightarrow \nu_{\tau}$ oscillations with parameters coming from
the {\it Up Through} sample.}
\end{figure}

\subsection{Low energy events}
The data sample for the analyses of low energy events
has been collected in $\sim 3$ yr of data
with the full detector.
The analysis of the {\it Internal Up} events is similar to the {\it Up
Through} analysis, with the additional requirement 
that the interaction vertex lies
inside the apparatus. From a full Monte Carlo using the Bartol neutrino 
flux \cite{Agrawal}, 
the $\nu$ cross-sections in Ref.~\cite{Lipari1} and the Lohmann \etal
muon energy loss \cite{Lohmann}, 
it is found that about 87$\%$ of the events are due to
$\nu_{\mu}$ CC interactions. The simulation has been performed in a 
large volume of rock (170 Kton) around the apparatus (5.3 Kton).
The theoretical uncertainty estimated for the calculation is about 25$\%$.
The preliminary estimated systematic error on
the acceptance and analysis requirements is $\sim 10\%$. After background 
subtraction (3 events) 85 events are selected as {\it Internal Up} events.

The {\it Internal Down} events and the {\it Upgoing Stopping} $\mu$s are
selected through topological cuts: the main requirement is the presence 
of a track crossing the bottom scintillator layer at least 
1 m from the detector walls. To reject ambiguous and/or wrongly
reconstructed tracks which survived the software cuts, real and simulated
events were randomly merged and visually scanned.
Three different samples have been selected according to the choice of the
minimum number of streamer plane hits. In Fig~\ref{fig4} we show the angular 
distribution of the 120 events (after the background subtraction of 5 events)
which cross at least 3 streamer planes 
(corresponding to $\sim 100$ g/cm$^{2}$) and the
angular distribution of the {\it Internal Up} events.

In Tab.~\ref{tab1} the total number of measured and expected events
(no-oscillation and $\nu_{\mu} \rightarrow \nu_{\tau}$ hypothesis) are
summarized for the two topologies.
The low energy events show a uniform (as a function of $\cos\theta$) deficit
of roughly a factor of 2 with respect to the expected
ones in the no-oscillation hypothesis. This result is in agreement with the
prediction based on the $\nu_{\mu} \rightarrow \nu_{\tau}$ oscillation 
hypothesis coming from the {\it Up Through} sample.
\begin{table}
\begin{center}
\footnotesize\rm
\caption{\label{tab1}
Low energy event summary. The predictions with oscillations are for 
$\sin^{2}2\theta = 1$ and $\Delta m^{2} = 0.0025$ eV$^{2}$.}
\begin{tabular}{llll} 
\topline
Topology & Data & MC (no-oscillation)& MC($\nu_{\mu} \rightarrow \nu_{\tau}$)
\\ 
\midline
{\it Up Through}     & 451& $612 \pm 144_{theor} \pm 37_{sys}$ & $431 \pm 
73_{theor} \pm 26_{sys}$ \\ 
{\it Internal Up}    &\085& $144 \pm \036_{theor} \pm 14_{sys}$& $\083 \pm
21_{theor} \pm \08_{sys}$\\ 
{\it Int. Down + UGS}& 120& $159 \pm \040_{theor} \pm 16_{sys}$& $123 \pm 
31 _{theor} \pm 12_{sys}$ \\ 
\bottomline
\end{tabular}
\end{center}
\end{table}

\section{The search for dark matter}
 
%
The existence of non-baryonic dark matter has been suggested due to 
many evidences indicating that the density of matter in the universe, 
$\Omega_{M}$ (in units of the critical density), is greater than 0.3. 
Theoretical prejudices
for a flat universe require the density of the universe to be $\Omega = 1$.
These indications and hypotheses together 
with the upper limits on luminous matter ($\Omega_{lum} \lesssim 0.01$)
\cite{Persic} 
and on baryonic mass from primordial 
baryonic nucleosynthesis (i.e. $0.005 \lesssim \Omega_{b} \lesssim 0.10$ at 
95$\%$ c.l. for $0.4 \lesssim h \lesssim 1$ \cite{Olive}) 
explain the need for non-baryonic dark matter.
It is a striking coincidence that if a weakly interacting massive 
dark matter particle (WIMP)
exists its abundance would correspond today nearly to the critical 
density due to the fact that the thermally averaged annihilation 
cross-section would be comparable to the typical weak interaction 
cross-section \cite{Jungman}.

If R-parity is conserved, supersymmetric (SUSY) theories would provide 
one of the most plausible candidate, the {\it Lightest Supersymmetric 
Particle} (LSP), which in most theories is the neutralino 
$\tilde{\chi}$, the lightest linear superposition
of gaugino and higgsino eigenstates. 
In Minimal Supersymmetric extensions of the Standard Model (MSSM)
including the GUT relation $M_{1} = M_{2} \tan^{2}\theta_{W}$, 
with $\theta_{W}$ the Weinberg angle, the
neutralino mass depends on the gaugino mass parameters $M_{1}$, 
on the higgsino mass parameter $\mu$ and on the ratio of the Higgs doublet 
vacuum expectation values $\tan\beta$.
Some other parameters must be determined in order to define the 
processes induced by neutralinos, such as $m_{A}$, the mass of the 
pseudoscalar Higgs boson, the squark and slepton common mass $m_{0}$ and 
the trilinear parameters of the $3^{rd}$ family with common value $A$.

The supersymmetric parameter space is constrained by collider searches
\cite{DPB} and lower limits on the neutralino mass are around 20-30 GeV.
These limits are model dependent and correlated to chargino limits.
In this framework, ``direct'' and ``indirect'' 
methods for detecting galactic halo
WIMPs performed with underground detectors can probe complementary regions
of the parameter space with respect to collider searches. Direct methods
employ low-background detectors (e.g. superconductors or scintillators)
to measure the energy deposited when a WIMP elastically scatters
from a nucleus; indirect methods measure neutrinos resulting from the
annihilation of WIMP pairs.

WIMPs in the halo can lose energy through elastic scattering in the
celestial bodies which they intercept and become gravitationally trapped
inside their core when their velocity falls below the escape velocity. 
As their density increases inside the core of the body, their annihilation 
rate increases until equilibrium is achieved between capture and annihilation
rate. High energy $\nu$s are eventually produced via the hadronization and/or
decay of the annihilation products (mostly fermion-antifermion pairs, weak
and Higgs bosons). Possible ``traps'' for WIMPs could be the Sun or the Earth.
Many calculations concerning the flux of neutrino induced upward-going muons
from neutralino annihilation in the Sun and the core of the Earth have been
made \cite{Jungman,Bottino,Bergstrom}.
Data on upward muons have been presented by several experiments
\cite{Baksan,Kamioka,IMB}.

\subsection{WIMP search with MACRO}

The aim of the search is to set flux limits (in absence of an evidence
for a signal) on the upward-going muon flux from the core of the Earth and of 
the Sun. Muon flux limits are evaluated (90$\%$ c.l.) as the ratio of 
the poissonian upper limit for the number of measured events and of expected  
background events due to atmospheric $\nu$s inside the 
search cone around the direction of the celestial body over the exposure
(live time times area seen by the source times detection efficiency).
 
\begin{figure}
\begin{center}
\epsfig{file=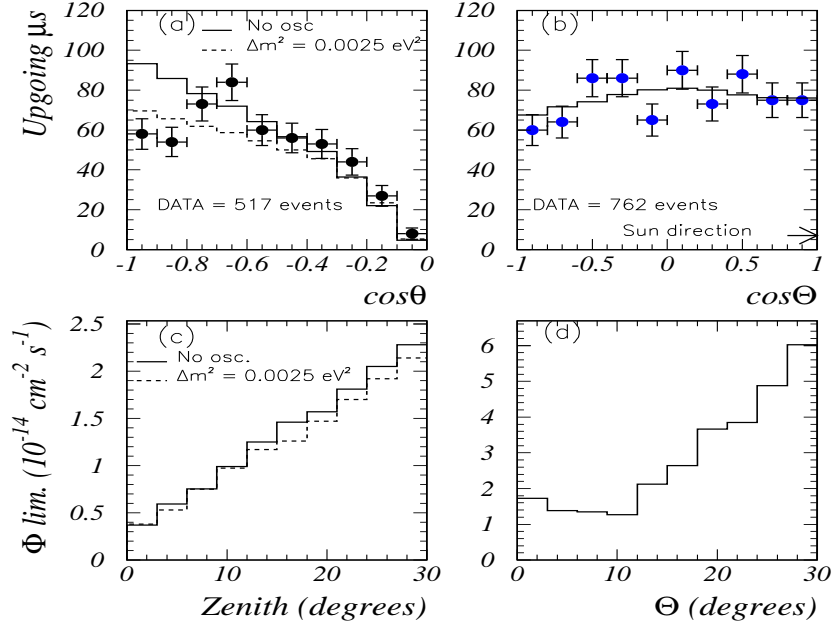,height=9.5cm,
width=12.cm}
\end{center}
\caption{\label{fig5} (a) Zenith distribution of measured
(circles) and expected (solid line) Up Through $\mu$. The expectation is 
normalized to the ratio data/MC for $\theta > 30^{\circ}$ equal to 0.85. 
(b) Distribution
of measured (circles) and expected (solid line) upward-going $\mu$s vs the 
cosine of the angular separation from the Sun direction. (c) Muon flux limits 
(90$\%$ c.l.) vs the zenith angle. In (a) and (c) the dotted line is for 
in the $\nu_{\mu} \rightarrow \nu_{\tau}$ oscillation hypothesis for the 
background of atmospheric $\nu$s with $\sin^{2}2\theta = 1$ and $\Delta m^{2} 
= 0.0025$ eV$^{2}$. In this case the normalization factor is 1.19. 
(d) Muon flux limits (90$\%$ c.l.) vs search angle around
the direction of the Sun.}
\end{figure}

For the search of WIMPs in the core of the Earth a larger sample of 517 
(collected up to March '98) upward 
throughgoing muons with respect to the analysis in section~\ref{Upth} 
has been used. After background subtraction the total resulting number
of events is $487 \pm 22_{stat}$ while the expected events are 
$652 \pm 111_{theor}$. 
As the background for this measurement overcomes the signal in the region
around the vertical of the apparatus, we set conservative flux limits for
WIMPs in the Earth assuming that the number of measured events
equals the expected one \cite{DPB}. 
The expected distribution has been normalized by a factor 0.85
(the ratio of measured to expected events with $\theta > 30^{\circ}$, which 
is outside the region interesting for the signal). The reason for this
normalization is that calculations on atmospheric $\nu$s are mainly affected by
a normalization theoretical uncertainty. The zenith angular distribution of 
measured and expected events is shown in Fig.~\ref{fig5}(a) and the 90$\%$ c.l.
muon flux limits for an average exposure of 2620 m$^2$ yr (the area decreases
of about 37 m$^2$ from 3$^{\circ}$ to 30$^{\circ}$)
are plotted in Fig.~\ref{fig5}(c) in search cones increasing 
from $3^{\circ}$ to $30^{\circ}$. Considering the results in 
Ref.~\cite{upmu98} (see section~\ref{Upth}), 
the hypothesis of $\nu_{\mu} \rightarrow \nu_{\tau}$
with $\sin^{2}2\theta = 1$ and $\Delta m^{2} = 0.0025$ eV$^{2}$ has been 
considered (dotted lines in Fig.~\ref{fig5}(a) and (c)). 

As for moving sources the background is less relevant than for steady sources,
in the case of the Sun 
we have used 762 events including {\it Internal
Up} events and events which cross less than 200 g/cm$^{2}$ through the 
absorber. The expectation has been obtained using the 
local coordinates of the measured upward-going events 
and randomly chosen detection times of downward-going muons 
measured during the data taking (times are needed in order to
evaluate equatorial coordinates). This procedure takes into account the 
drifts of detection efficiency in time. Fig.~\ref{fig5}(b) shows
the distribution of the 762 events as a function of their angular
separation with respect to the Sun direction. In this distribution
events detected during night fall towards -1. In Fig.~\ref{fig5}(d)
the muon flux limits are shown as a function of 10 arbitrary search cones
around the direction of the Sun for an exposure of $\sim 890$ m$^2$ yr.   
Some of the limits are summarized in Tab.~\ref{tab2}. The minimum muon energies
have been chosen where the acceptance of the detector becomes independent 
on energy.  
It has an higher value for the Sun because tracks from the Sun direction
are more slanted than the ones at the vertical of the detector.
\begin{table}
\begin{center}
\footnotesize\rm
\caption{\label{tab2} Selected and expected events and $90\%$ c.l.  
muon flux limits for some of the 10 half-cones chosen around the core of the
Earth (exposure $\sim$ 2620 m$^{2}$ yr, $E_{\mu ,\ min} > 1.5$ GeV) 
and the Sun (exposure $\sim$  890 m$^{2}$ yr, $E_{\mu ,\ min} > 2$ GeV) 
direction.}
\begin{tabular}{ccccccc} 
\topline
\multicolumn{1}{c}{ }&
\multicolumn{3}{c}{EARTH}&
\multicolumn{3}{c}{SUN} \\ \hline
\multicolumn{1}{c}{Cone}  
&\multicolumn{1}{c}{Data} 
&\multicolumn{1}{c}{Backgr.}
&\multicolumn{1}{c}{Flux Limit} 
& \multicolumn{1}{c}{Data} 
& \multicolumn{1}{c}{Backgr.} 
&\multicolumn{1}{c}{Flux Limit} \\ 
     &       &            &(cm$^{-2}$ s$^{-1}$)& &  
& (cm$^{-2}$ s$^{-1}$)\\ \hline
$30^{\circ}$ &76& 123.6& 2.28 $\times 10^{-14}$&56 & 51.1& 
6.02 $\times 10^{-14}$\\
$24^{\circ}$ &52& 81.2 & 1.81 $\times 10^{-14}$&33 & 33.0& 
3.85 $\times 10^{-14}$ \\
$15^{\circ}$ &24& 32.5 & 1.25 $\times 10^{-14}$&11 & 13.0 & 
2.12 $\times 10^{-14}$\\
$9^{\circ}$ &10& 11.9 & 7.54 $\times 10^{-15}$ & 3 &  4.6 & 
1.35 $\times 10^{-14}$\\
$3^{\circ}$  &0 &  1.3 & 3.72 $\times 10^{-15}$& 2 &  0.5 & 
1.73 $\times 10^{-14}$\\
\bottomline
\end{tabular}
\end{center}
\end{table}

\subsection{Upper limits for neutralinos}

In the previous section flux limits have been calculated
in arbitrary search cones, so that any WIMP model can be constrained
by MACRO data. Here we assume that WIMPs are neutralinos and hence we
calculate flux limits in the search cones which collect 90$\%$ of the
signal from neutralino annihilation. The muons from
$\tilde{\chi}$ 
annihilation are distributed around the neutrino direction because
of the $\nu-\mu$ angle due to the $\nu$ CC interaction in the Earth and
to the multiple scattering of muons along the path from the $\nu$ interaction
point to the detector. The $\nu-\mu$ angle distribution depends on the
$\nu$ energy spectrum and hence in principle on the details of the final 
states of the annihilation process. However, the relevant parameter is
$m_{\tilde{\chi}}$. The branching ratios into the final states depend on the
supersymmetric model. We estimate that the higher variation of
the width of the $\nu-\mu$ angle is of the order of 17$\%$
when an extreme model (in which only one channel is allowed)
is used instead of a more plausible model (as the one used here).

The shapes of the upward-going muon signal has been calculated using $\nu$ 
fluxes from neutralino annihilation in the Sun and in the Earth given 
by Bottino \etal \cite{Bottino}, the $\nu$ CC cross sections in 
Ref.~\cite{Lipari1}, the muon propagation in Ref.~\cite{Lohmann} and
taking into account the angular resolution of the detector. We have 
considered the $\tilde{\chi}$ distribution in the core of the Earth
\cite{Bottino}.	
\begin{figure}
\begin{tabular}{cc}
\epsfig{file=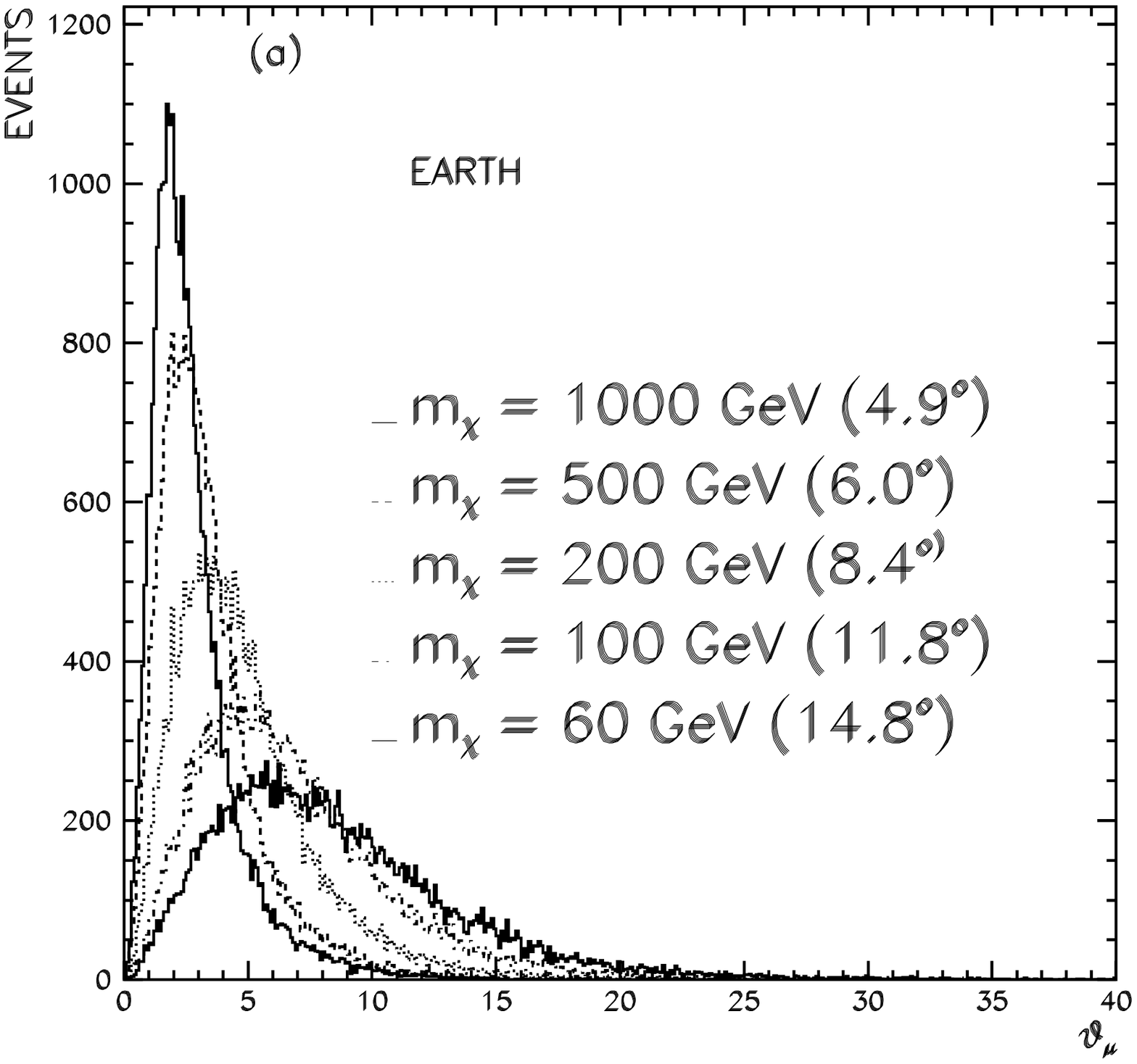,height=6.cm,
width=5.5cm}&
\epsfig{file=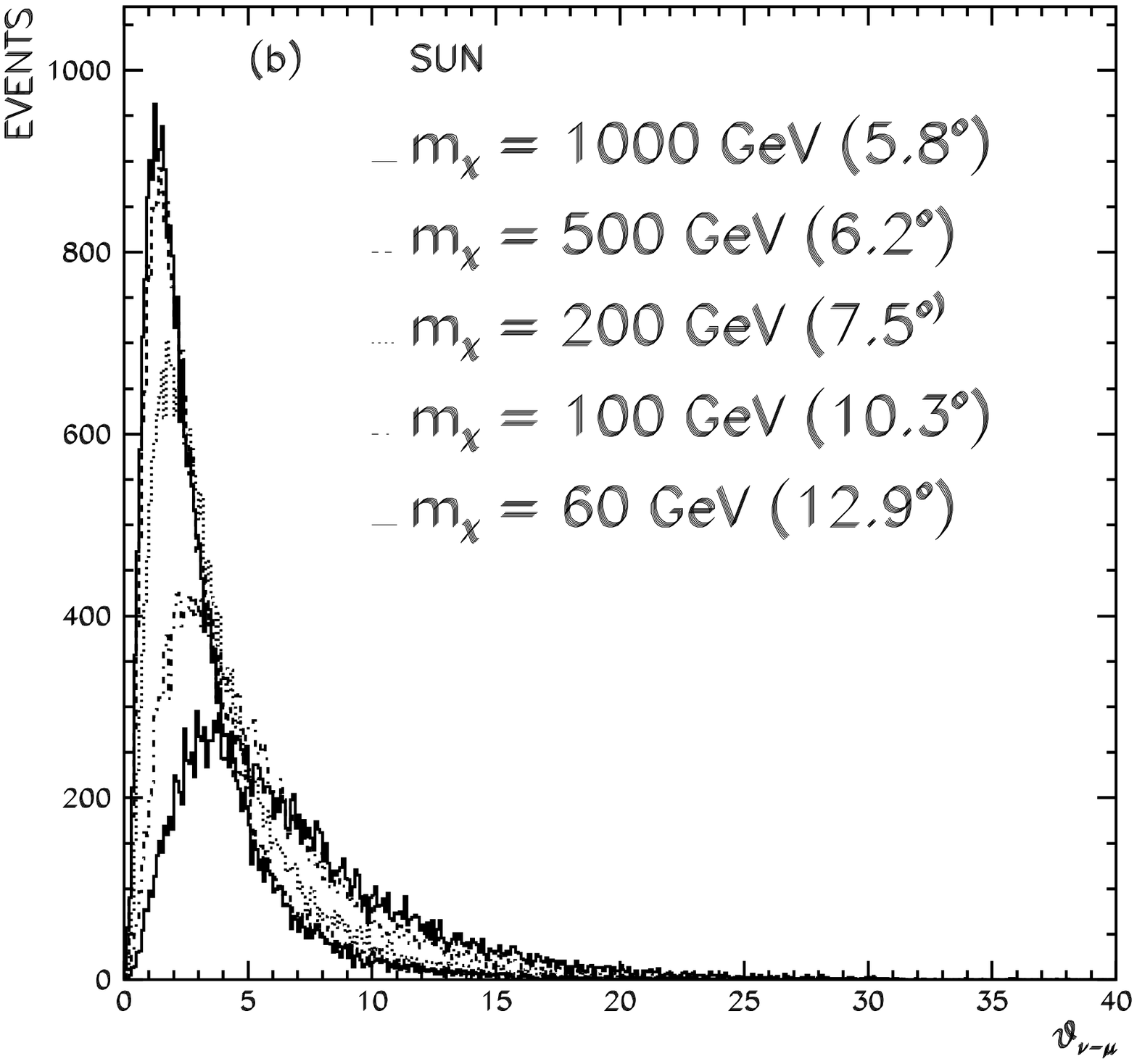,height=6.cm,
width=5.5cm}
\end{tabular}
\caption{\label{fig6}
(a) Nadir angle distribution of $\mu$s from neutralino annihilation
inside the Earth for different masses. (b) $\nu-\mu$ angle distribution
for neutralino annihilation in the Sun and different masses.
In (a) and (b) the angular ranges including 90$\%$ of the signal are 
indicated.}
\end{figure}
The 90$\%$ c.l. flux limits
have been calculated in neutralino mass dependent 
cones which collect 90$\%$ of the signal shown in Fig.~\ref{fig6}
for the Earth and the Sun for $m_{\tilde{\chi}} = 60, 100, 200, 500, 1000$ GeV.
In Fig.~\ref{fig7} 
the experimental limits are superimposed to the fluxes of upward-going
muons from Bottino \etal \cite{Bottino} calculation as a function of 
$m_{\tilde{\chi}}$
for a muon minimum energy of 1 GeV (the dependence on the energy of MACRO 
acceptance has been taken into account in the low energy region).
The fluxes are calculated
varying the model parameters (each model is represented by a dot in 
Fig.~\ref{fig7}) in experimentally allowed ranges:
$10 \le \left| \mu \right| \le 500$ GeV, $10 \le M_{2} \le 500$ GeV,
$1.01 \le \tan\beta \le 50$, $65 \le m_{A} \le 500$ GeV, 
$150 \le m_{0} \le 500$ GeV, $-3 \le A \le 3$, where $m_{A}$ is the 
mass of the pseudoscalar Higgs, $m_{0}$ is the common soft mass of all the 
sfermions and squarks, $A$ is the common value of the
trilinear coupling in the superpotential for the third family,
with central values of allowed interval for cosmological 
parameters (rms velocity of $\tilde{\chi}$ in the halo = 270 km s$^{-1}$, 
$\tilde{\chi}$ escape velocity in the halo = 650 km s$^{-1}$,
velocity of the Sun around the galactic center = 232 km s$^{-1}$, 
local dark matter density $\rho_{loc} = 0.5$ GeV 
cm$^{-3}$ and minimal value for rescaling the neutralino relic abundance
$\Omega h^{2}_{min}$ = 0.03). In fact,
Bottino {\sl et al.} assume: 
\begin{eqnarray}
\begin{array}{cc}
{\rm if}\; \Omega h^{2}_{\tilde{\chi}} > (\Omega h^{2})_{min} & \rho_{\tilde{
\chi}} = 
\rho_{loc} \\
{\rm if}\; \Omega h^{2}_{\tilde{\chi}} < (\Omega h^{2})_{min} & \rho_{\tilde{
\chi}} = 
\rho_{loc} \times  \Omega h^{2}_{\tilde{\chi}} / (\Omega h^{2})_{min} \, .
\end{array}
\end{eqnarray}
where $(\Omega h^{2})_{min}$ = 0.03.
The fluxes lying above the experimental limit are ruled
out as possible SUSY models by this measurement.
Fig.~\ref{fig7} indicates that indirect search can have good prospects
particularly at high $\tilde{\chi}$ masses.

\section{Conclusions}
The upward throughgoing muon MACRO data set is
in favor of a $\nu_{\mu} \rightarrow \nu_{\tau}$ hypothesis with
$\Delta m^{2} \sim 0.0025$ eV$^{2}$ and maximum mixing with maximum
probability of 17$\%$ against the 0.1$\%$ probability obtained for
the no-oscillation hypothesis. These oscillation parameters 
are very close to the ones preferred by the SuperKamiokande experiment
\cite{SK}.
The hypothesis found is in agreement with the first results concerning the
MACRO low energy neutrino events. However, the study of the shape of the
upward throughgoing muon angular distribution 
gives a maximum probability of only 4.6$\%$. This
could be due to a statistical fluctuation or to some hidden physics. Having
performed numerous systematic checks, we exclude any effect on the anomalous
zenith distribution due to the detector.

A search for a WIMP signal has been performed using throughgoing upward
$\mu$s coming from the Earth core and from the Sun.
At this time we have observed no signal for this signature and 
we have calculated upper limits for the flux of upward $\mu$s from WIMPs. 
These flux limits have been compared with predictions of 
SUSY models.  
Our data have begun to rule out significant portions of the parameter 
space for WIMPs annihilating in the Earth and in the Sun
and currently limits for the Earth are the most stringent limits of all those
from ``indirect'' experiments \cite{Baksan,Kamioka}.

\normalsize
\begin{figure}
\begin{tabular}{cc}
\epsfig{file=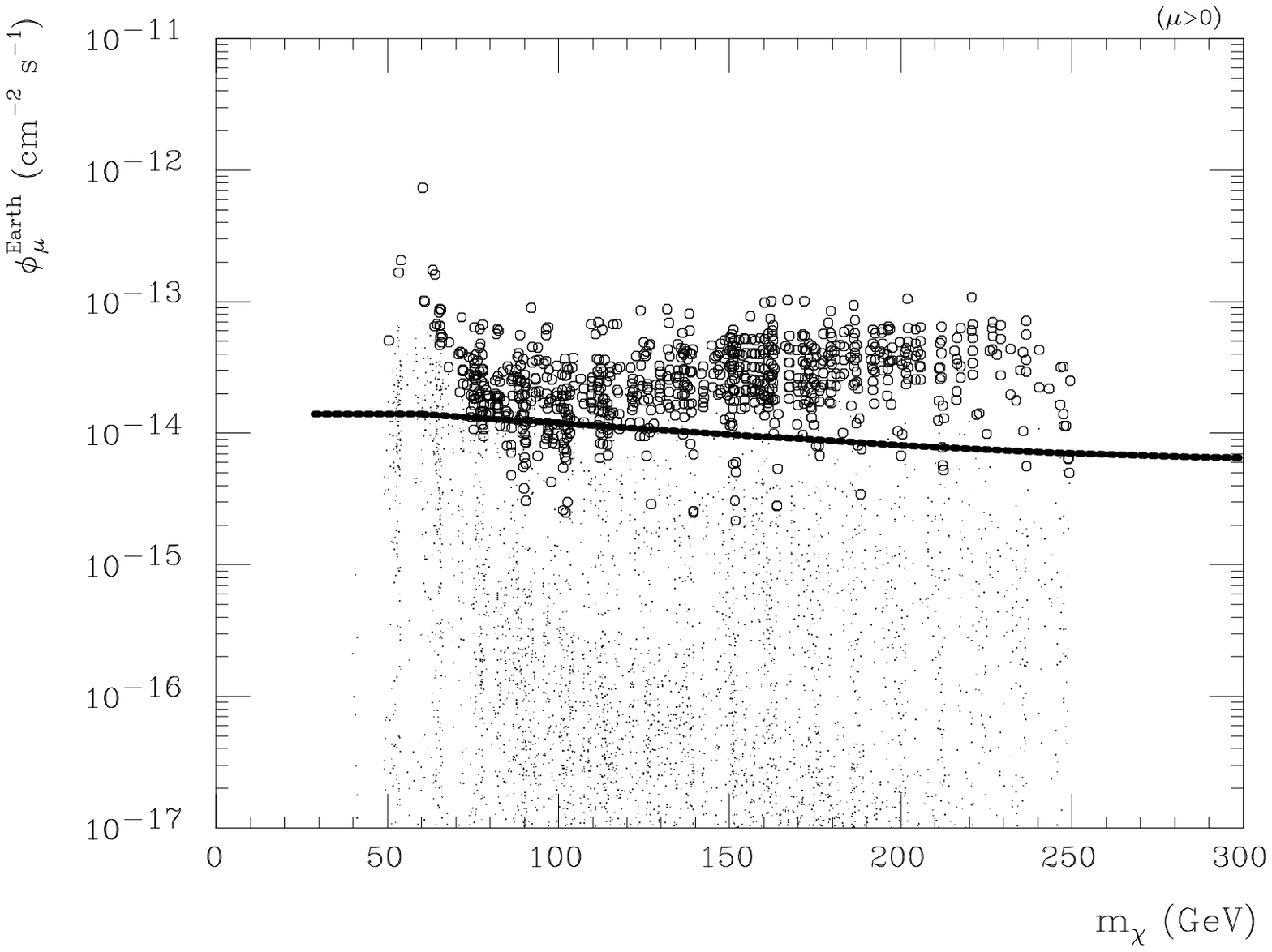,height=9.cm,
width=5.5cm}&
\epsfig{file=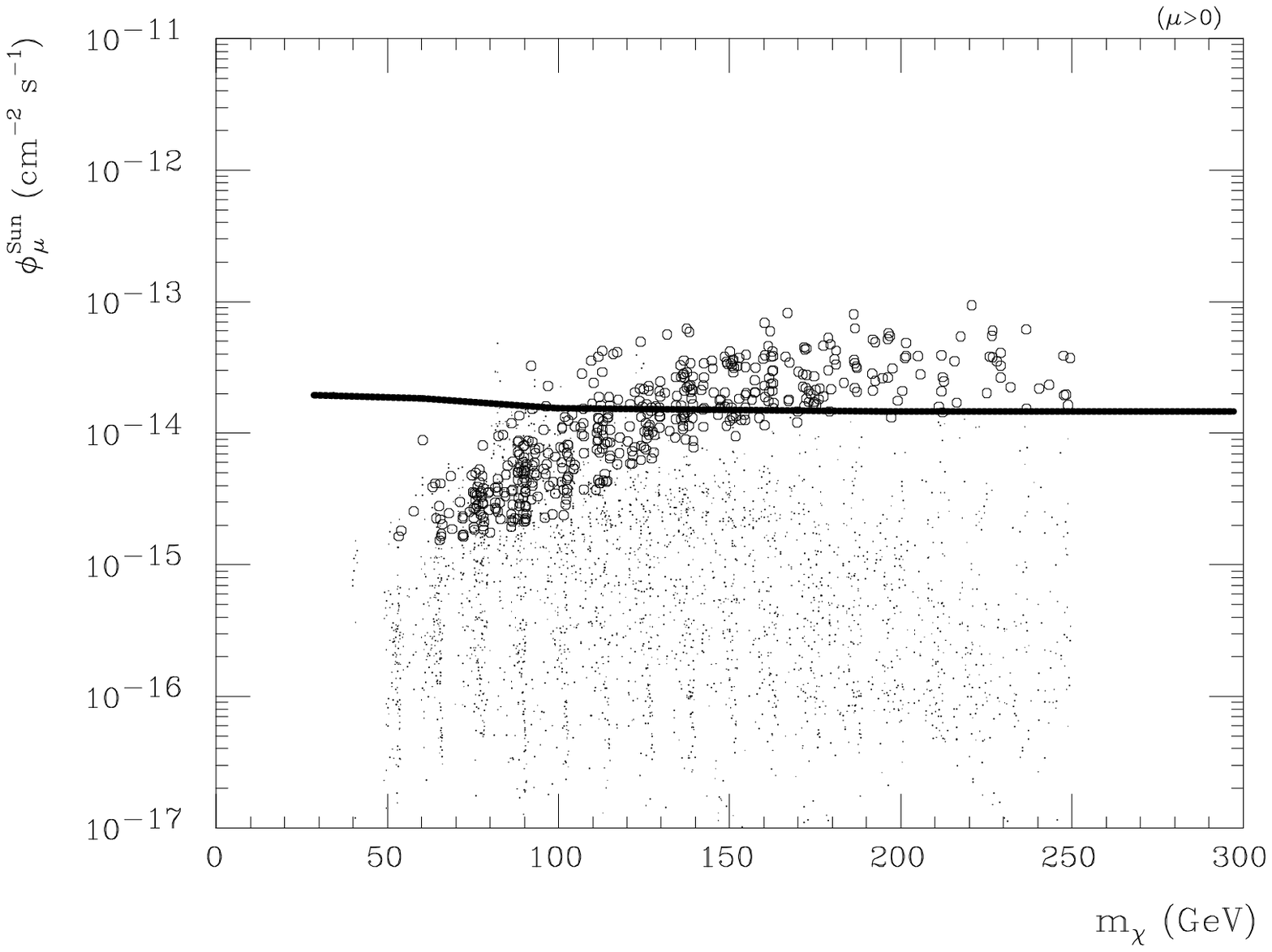,height=9.cm,
width=5.5cm}
\end{tabular}
\caption{\label{fig7}
Upgoing $\mu$ flux vs $m_{\tilde{\chi}}$ for $E_{\mu}^{th} >$ 1 GeV
from the Earth (plot on the left) and the Sun (plot on the right)
\protect\cite{Bottino}. Solid line: MACRO flux limit (90$\%$ signal). 
Open circles: region excluded by direct measurements
(particularly the NaI DAMA experiment \protect\cite{DAMA}) for 
local dark matter density 0.5 GeV cm$^{-3}$.}
\end{figure}
\end{document}